\documentclass[prb, 10pt,twocolumn,letterpaper,superscriptaddress,floatfix,nofootinbib,notitlepage]{revtex4-1}

\usepackage{textcomp}
\usepackage{amsmath}
\usepackage{amsfonts}
\usepackage{amssymb}
\usepackage{graphicx}
\usepackage{siunitx}
\usepackage{caption}
\usepackage{setspace}
\usepackage{xcolor}
\setcitestyle{super}
\usepackage[colorlinks, citecolor={blue}, linkcolor={black}]{hyperref}

\begin{document}
\raggedbottom

\title{
Forbidden second harmonics in centrosymmetric bilayer crystals
}

\author{Haoning Tang$^+$}
\email{haoning@mit.edu}
\affiliation{Department of Electrical Engineering and Computer Science, University of California at Berkeley, Berkeley, CA 94720, USA}
\affiliation{Department of Mechanical Engineering, Massachusetts Institute of Technology, Cambridge, MA 02139, USA}

\author{Zhitong Ding$^+$}
\affiliation{Department of Electrical Engineering and Computer Science, University of California at Berkeley, Berkeley, CA 94720, USA}

\author{Tianyi Ruan}
\affiliation{Department of Electrical Engineering and Computer Science, University of California at Berkeley, Berkeley, CA 94720, USA}

\author{Zeyu Hao}
\affiliation{Department of Electrical Engineering and Computer Science, University of California at Berkeley, Berkeley, CA 94720, USA}
\affiliation{Materials Sciences Division, Lawrence Berkeley National Laboratory, Berkeley, CA, USA}

\author{Kenji Watanabe}
\author{Takashi Taniguchi}
\affiliation{National Institute for Materials Science, Namiki 1-1, Tsukuba, Ibaraki 305-0044, Japan}

\author{Haozhe Wang}
\affiliation{Department of Electrical and Computer Engineering, Duke University, Durham, NC 27708, USA}

\author{Ali Javey}
\affiliation{Department of Electrical Engineering and Computer Science, University of California at Berkeley, Berkeley, CA 94720, USA}
\affiliation{Materials Sciences Division, Lawrence Berkeley National Laboratory, Berkeley, CA, USA}

\author{Feng Wang}
\affiliation{Department of Physics, University of California at Berkeley, Berkeley, CA 94720, USA}
\affiliation{Materials Sciences Division, Lawrence Berkeley National Laboratory, Berkeley, CA, USA}

\author{Yuan Cao}
\email{caoyuan@berkeley.edu}
\affiliation{Department of Electrical Engineering and Computer Science, University of California at Berkeley, Berkeley, CA 94720, USA}
\affiliation{Department of Physics, University of California at Berkeley, Berkeley, CA 94720, USA}
\affiliation{Materials Sciences Division, Lawrence Berkeley National Laboratory, Berkeley, CA, USA}

\begin{abstract}
Optical spectroscopy based on second-order nonlinearity is a critical technique for characterizing two-dimensional (2D) crystals\cite{kumar_second_2013,malard_observation_2013,li_probing_2013,kim_stacking_2013} as well as bioimaging and quantum optics \cite{campagnola_second-harmonic_2003,zhang_spontaneous_2021}. It is generally believed that second-harmonic generation (SHG) in centrosymmetric crystals, such as graphene and other bilayer 2D crystals\cite{zhang_second_2020}, is negligible without externally breaking the inversion symmetry. Here, we show that with a new homodyne detection technique, we can apparently circumvent this symmetry-imposed constraint and observe robust SHG in pristine centrosymmetric crystals, without any symmetry-breaking field. With its exceptional sensitivity, we resolve polarization-resolved SHG in bilayer hexagonal boron nitride (h-BN), bilayer 2H-WSe\textsubscript{2}, and remarkably, Bernal-stacked bilayer graphene, allowing us to unambiguously identify the crystallographic orientation in these crystals via SHG for the first time. We also demonstrate that the new technique can be used to non-invasively detect uniaxial strain and optical geometric phase in these crystals. The observed SHG in our experiments is attributed to second-order nonlinearity in the quadrupole channel, which is controlled by the presence of the $C_2$ symmetry instead of the inversion symmetry. Our new technique expands the capability of nonlinear optical spectroscopy to encompass a large class of centrosymmetric materials that could never be measured before, and can be used for quantum sensing of moiré materials and twisted epitaxial films. 
\end{abstract}

\maketitle

\section{Main}
Two-dimensional crystals such as graphene and transitional metal dichalcogenides (TMDs) are host to crystalline surfaces with the highest quality and tunability. The advent of moiré engineering on these materials has further pioneered novel approaches for creating new quantum materials and devices\cite{andrei_marvels_2021}, leading to the creation of new exotic correlated electronic phases such as unconventional superconductivity\cite{cao_unconventional_2018}, fractional quantum anomalous Hall states\cite{cai_signatures_2023}, and Wigner crystals\cite{regan_mott_2020,xu_correlated_2020}. A crucial aspect in all of these experiments is the precise control of the crystallographic orientation of the underlying 2D crystals. It is therefore highly desirable to develop a universal, quick, and non-contact approach to characterize the crystallographic orientation of any 2D crystal with no ambiguity. 

For this purpose, nonlinear optical methods based on second-harmonic generation (SHG) are frequently employed\cite{kumar_second_2013,malard_observation_2013,li_probing_2013,kim_stacking_2013}, but not without limitations. In a typical SHG measurement, a femtosecond laser at wavelength $\lambda_0$ (fundamental wavelength) is focused at the sample, and the reflected (or transmitted) light at $\lambda_0/2$ (second-harmonic wavelength) is measured. The second-harmonic signal light is often projected to the same polarization plane as the fundamental light by an analyzer to achieve a parallel input/output configuration. A 2D crystal with hexagonal lattice will exhibit an SHG intensity proportional to $\cos^23\theta$ if it breaks the inversion symmetry, where $\theta$ is the angle between the armchair direction of the crystal and the polarization plane. This results in polarization-dependent SHG intensity that is periodic in $60^\circ$\cite{boyd_nonlinear_2020}. However, two limitations with conventional SHG are immediately apparent: (i) traditional SHG cannot be used to determine the crystallographic orientation of centrosymmetric materials, as their second-order nonlinear tensor $\chi^{(2)}$ vanishes. These include some of the most important 2D crystals used in moiré engineering, such as monolayer and Bernal bilayer graphene, as well as bilayers of h-BN and 2H-phase transitional metal dichalcogenides (TMDs). (ii) Even for noncentrosymmetric materials (e.g. odd-layer h-BN and 2H-TMDs) that do exhibit nonvanishing $\chi^{(2)}$, an inconvenient ambiguity remains: conventional SHG cannot distinguish between 2D crystals rotated at $\theta$ and $\theta+60^\circ$. These are nonequivalent crystals, and moiré materials created from one versus the other behave very differently\cite{mak_semiconductor_2022}. These inherent limitations render traditional SHG an inadequate approach for deterministic production of moiré heterostructures.

In this article, we introduce a powerful and yet simple technique that simultaneously circumvents both of these inherent limitations in traditional SHG. With this ultra-sensitive technique, we are able to unambiguously measure the second-order optical response and extract the crystallographic orientations of any layers of h-BN, 2H-TMDs, or graphene (except monolayer graphene), in particular bilayers of them that possess inversion symmetry, for the first time. We point out that the observed SHG in centrosymmetric crystals is quadrupolar in nature and is constrained by the $C_2$ symmetry rather than the inversion symmetry. We also demonstrate strain and geometric phase sensing using the new technique. These results open the doors towards precise optical sensing, characterization, and manipulation of a broad range of crystalline materials than previously thought to be impossible for SHG. The new technique is especially useful for future `twistronic' studies that require precise alignment between dissimilar materials, and might also be utilized for quantum sensing of $C_2$-broken correlated phases in moiré materials. 

\begin{figure*}[!h]
    \centering
    \includegraphics[width=0.9\linewidth]{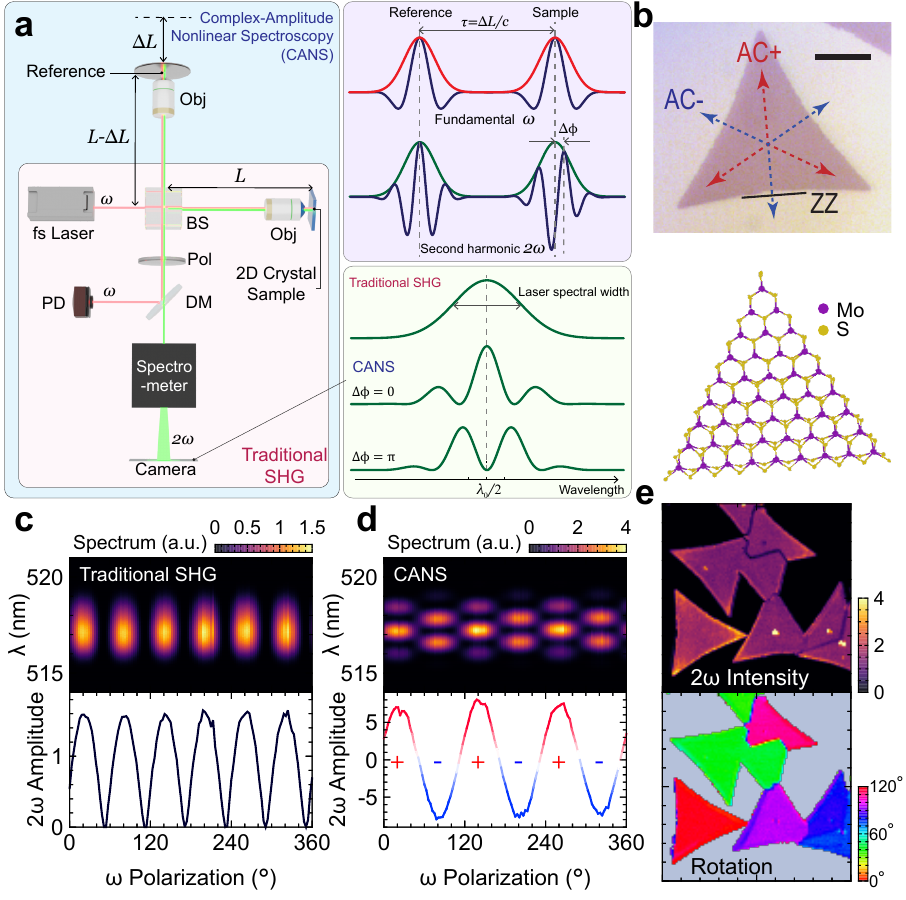}
    \caption{\small\textbf{An ultrasensitive phase-resolved second-harmonic spectroscopy.} (a) Illustration of CANS technique in comparison to traditional SHG spectroscopy. In CANS, coherent SHG waves from the reference and the sample interfere at the beam splitter (BS) to create a spectrum that contains the phase-resolved SHG amplitude from the sample. The upper-right panel shows the time domain pulses from the reference and the sample, at the fundamental (red) and second-harmonic (green) wavelengths. The lower-right panel shows the spectrum near $\lambda_0/2$, in a traditional SHG measurement, as well as in CANS measurements, when a phase difference of $\Delta\phi=0$ or $\pi$ is present between the reference SHG and sample SHG waves. (b) In a monolayer MoS\textsubscript{2} flake, the armchair directions from Mo to S (AC+) are nonequivalent with the armchair directions from S to Mo (AC-). Scale bar is \SI{5}{\micro\meter}. (c) Polarization-dependent spectrum and SHG amplitude (square root of intensity) in a traditional SHG measurement. The six polarizations that yields maximum SHG signal are the armchair directions. However, the AC+ and AC- directions cannot be distinguished. (d) Spectrum and SHG amplitude in a CANS measurement. The phase information of the SHG wave is encoded into the interference pattern in the wavelength domain. The phase-resolved SHG amplitude extracted from CANS clearly differentiates AC+ and AC- directions. (e) SHG intensity map and crystallographic rotation map of a region with multiple overlapping MoS\textsubscript{2} monolayer flakes. By using CANS, the crystallographic orientation of a 2D crystal can be determined unambiguously, respecting the underlying $C_3$ symmetry. Both maps are $100\times$\SI{100}{\micro\meter\squared}.}
    \label{fig:fig1}
\end{figure*}

\section{Complex-amplitude nonlinear spectroscopy (CANS)}

In a nonlinear crystal, the SHG is produced by the second-order electric polarization density $\mathbf{P}^{2\omega}=\chi^{(2)} \mathbf{E}^\omega\mathbf{E}^\omega$, where $\mathbf{E}^\omega$ is the incident fundamental electrical field and $\omega$ is the optical angular frequency. It is a fully coherent process, in which $\mathbf{P}^{2\omega}$ oscillates in full synchronization with $\mathbf{E}^\omega$. In a traditional SHG measurement in parallel input/output configuration, one measures the intensity (square of amplitude) of the projected second-harmonic optical field, which we define as $I_{SHG,||}$. The polarization direction that yields the maximum $I_{SHG,||}$ corresponds to the armchair directions in a hexagonal 2D crystal \cite{kumar_second_2013,malard_observation_2013,li_probing_2013,kim_stacking_2013}. Since no phase information is retained in this measurement, traditional SHG does not resolve the \emph{sign} of the SHG amplitude. 

To see why this lack of phase information leads to an ambiguity in the determination of crystallographic orientation, one could consider a monolayer MoS\textsubscript{2} crystal as an example, as shown in Fig. 1b. While traditional SHG can differentiate between the armchair (AC) directions (maximum intensity) versus the zigzag (ZZ) directions (minimum intensity), one cannot tell between the directions along Mo$\rightarrow$S bonds (AC+) versus those along S$\rightarrow$Mo bonds (AC-). These two sets of directions are nonequivalent, and must be differentiated for reproducible fabrication of moiré heterostructures involving two different crystals. In terms of the rotational angle, intensity measurements can only resolve the angle of the crystallographic axis modulus $60^\circ$, instead of modulus $120^\circ$, which is prescribed by the $C_3$ rotational symmetry of MoS\textsubscript{2}. 
 
A phase-sensitive (interferometric) measurement is thus required to obtain the full crystallographic information encoded in $\mathbf{P}^{2\omega}$\cite{stolle_phase_1996}. To resolve the phase of the SHG from the sample under test, one needs a reference sample from which the phase relationship between the fundamental and second-harmonic wave is known (e.g. GaAs). By interfering the sample SHG wave with the reference SHG wave, one could obtain their phase difference and reconstruct the phase of the $\mathbf{P}^{2\omega}$\cite{kim_second-harmonic_2020,noauthor_phase-referenced_nodate,wang_contrast-enhanced_2024,shen_phase-sensitive_2013}. For this purpose, we build a nonlinear Michelson interferometer to automatically achieve the phase referencing, which is illustrated in Fig. 1a. On top of a traditional SHG setup, the fundamental laser beam is split at a beam splitter (BS) and focused onto the reference and the sample with identical microscope objectives. The reflected fundamental light from the sample and the reference are recombined at the BS and monitored with a photodiode (PD) in a closed-loop manner to ensure that the optical path difference between the two arms of the interferometer $\Delta L=n\lambda_0$, where $n$ is an integer (see SI for full setup). Therefore, $\mathbf{E}^\omega$ is guaranteed to arrive at the reference and the sample precisely in-phase. In the meantime, the reflected SHG from the sample and the reference are also combined at the BS (green beams in Fig. 1a). 

The fundamental and SHG waves combined at the BS in the time domain resemble the illustration on the upper-right panel in Fig. 1a. In the time domain, each femtosecond pulse from the pump laser generates two pulses, one from the reference and one from the sample, separated by a time delay of $\Delta L/c$, where $c$ is the speed of light. While the two fundamental pulses are guaranteed to have identical time-domain profiles (their peak heights could differ), the two SHG pulses could have a finite phase difference $\Delta\phi$  due to the inherent phase difference between the SHG of the reference and of the sample. $\Delta\phi$ thus directly probes the phase of $\mathbf{P}^{2\omega}$ from the sample. 

This technique, which we name as Complex-Amplitude Nonlinear Spectroscopy (CANS), essentially uses the fundamental pulses as a precise gauge to time the phase difference between the two SHG pulses $\Delta\phi$\cite{chang_relative_1965}. To facilitate phase extraction, the SHG signal is separated from the fundamental and analyzed with a spectrograph equipped with a camera. The double pulse in the time domain generates an interference pattern in the wavelength (frequency) domain according to its Fourier transform, much in analogy to a Young's double-slit experiment, as illustrated in the lower-right panel of Fig. 1a. This interference pattern is convoluted with the up-converted laser spectral width. If a nonzero $\Delta\phi$ exists between the two SHG pulses, there will be a phase shift in the interference pattern. Using a simple wavelet analysis on the spectrum near $\lambda_0/2$, both SHG amplitude (instead of intensity) and SHG phase ($\Delta \phi$) can be readily obtained. (see SI for data analysis procedure). 

As a benchmark, we perform both traditional SHG and CANS measurement on the same monolayer MoS\textsubscript{2} single-crystal (Fig. 2b), which has armchair and zigzag directions that are unambiguously determinable from its triangular shape. While traditional SHG spectroscopy (Fig. 1c) only shows the standard six-fold pattern in the polarization-resolved SHG measurement, the interference pattern in CANS reveals additional sign information, allowing direct correlation of positive CANS signal to the AC+ direction and negative CANS signal to the AC- direction. It is evident that the polarization dependence of CANS respects the underlying $C_3$ symmetry of monolayer MoS\textsubscript{2} and thus does not have any ambiguity. We can also measure the CANS map of multiple MoS\textsubscript{2} flakes by scanning the laser spot over the sample, as shown in Fig. 1e. From the complex-amplitude SHG data, we can directly extract the rotation angle of each individual region with high precision\cite{mueller_full_nodate}.

Another key advantage of CANS is its exceptional sensitivity. In traditional SHG measurements, the measured signal (SHG intensity) is proportional to $|P^{2\omega}_\mathrm{sam}|^2$ (we use subscripts `sam' and `ref' to refer to SHG contribution from the sample and the reference, respectively). On the other hand, the SHG amplitude signal in a CANS measurement is proportional to $P_\mathrm{ref}^{2\omega}\cdot P_\mathrm{sam}^{2\omega}$, which fundamentally classifies CANS as a homodyning detection technique\cite{thiansathaporn_homodyne_1995,dadap_homodyne_1999}. For samples with weak $P_\mathrm{sam}$, CANS is orders of magnitude more sensitive than the traditional SHG, making it an ideal technique for detecting weak SHG signals where the signal-to-noise ratio is limited by the background or the sensor\cite{thiansathaporn_homodyne_1995}. We will show next how this could be exploited to observe SHG in materials that were long thought to be lacking of SHG response.

\begin{figure*}[!ht]
    \centering
    \includegraphics[width=0.9\linewidth]{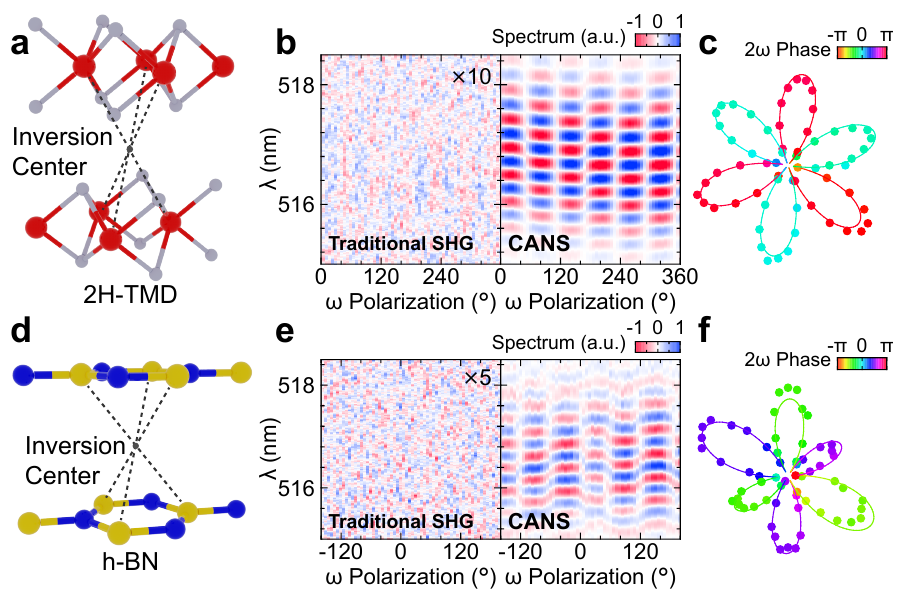}
    \caption{\textbf{Second harmonic generation from centrosymmetric crystals}. (a) Crystal structure of bilayer 2H-TMDs, showing its inversion center. (b) While traditional SHG can hardly detect any SHG signal from a bilayer 2H-WS\textsubscript{2}, CANS can clearly detect an SHG interference pattern. See SI for background removal procedure for both data. The traditional SHG data is scaled by 10 times. (c) The complex-valued SHG amplitude extracted from the CANS measurement in (b). The SHG amplitude is represented by the radial distance (arbitrary units), and the SHG phase is represented by color (from $-\pi$ to $\pi$). The solid curve is a fitting curve to the data points (see SI for the fitting function). (d) Crystal structure of bilayer h-BN and its inversion center. (e-f) Same measurements as in (b-c) for bilayer h-BN. Traditional SHG in (e) is scaled by 5 times. The distortion seen in (f) is attributed to uniaxial strain in the crystal. }
    \label{fig:fig2}
\end{figure*}

\section{SHG from centrosymmetric crystals}

Centrosymmetric crystals have vanishing $\chi^{(2)}$ because $\chi^{(2)}$ is odd under inversion. For this reason, while 2D crystals such as monolayer TMDs and h-BN break inversion symmetry and exhibit a strong SHG response, inversion symmetry is restored in bilayer 2H-TMDs (Fig. 2a) and bilayer h-BN (Fig. 2d), and these crystals show negligible SHG response in their pristine states\cite{li_probing_2013}. This symmetry-imposed constraint makes it impossible to characterize the crystallographic properties of these bilayer crystals using SHG\cite{boyd_nonlinear_2020,zhang_second_2020}, unless the inversion symmetry is externally broken by various methods, such as strong substrate effect \cite{dean_second_2009, tom_second-harmonic_1983}, van der Waals stacking\cite{yao_enhanced_2021,kim_stacking_2013,hsu_second_2014,zhang_emergent_2023}, application of a dc electric field or dc bias current \cite{an_enhanced_2013,yu_charge-induced_2015,klein_electric-field_2017,zhang_doping-induced_2019,lee_electrical_2016}, magnetic ordering \cite{ahn_electric_2024}, excitonic polarization \cite{shree_interlayer_2021}, strain gradient\cite{lu_strong_2023,zhang_strain-driven_2025}, among others. These approaches are either only applicable to certain classes of 2D crystals, or require special sample preparation that renders them unsuitable as non-contact characterization techniques. 

Here we demonstrate that the exceptional sensitivity of CANS allows one to apparently `evade' this symmetry constraint, enabling the universal and robust observation of SHG in centrosymmetric bilayers. SHG measured via CANS does not depend on any particular material traits (e.g. excitonic resonances), and does not require any external symmetry breaking fields (e.g. doping, gating, current biasing, or straining). Figure 2b compares the traditional SHG spectrum and CANS for a bilayer WSe\textsubscript{2} sample. These two measurements were taken in otherwise identical conditions (pump laser power, exposure time, etc.), but one with the reference signal blocked (traditional SHG) and one with it unblocked (CANS). We use normal incidence in all of our measurements to maximize the spatial resolution. The traditional SHG spectrum hardly reveal any SHG signal for this sample, as one would expect based on its inversion symmetry. To our surprise, a strong SHG signal is present in the CANS measurement, manifested as an interference pattern in the wavelength domain. We extract the polarization-resolved complex-valued SHG amplitude from this interference and show it in the radial plot in Fig. 2c, where the amplitude (modulus) and phase are represented by the radial distance and color, respectively. The polarization dependence of the CANS signal has a clear $120^\circ$ periodicity, and is in fact similar to that of monolayer 2H-TMD (Fig. 1d) \cite{kumar_second_2013,malard_observation_2013,li_probing_2013,kim_stacking_2013}.

Optically generated exciton-polaritons in bilayer 2H-TMDs are known to break the inversion symmetry in certain scenarios and can lead to enhanced SHG\cite{shree_interlayer_2021}. To rule out this scenario and to demonstrate that our observation of SHG in centrosymmetric crystals is universal, we performed the same measurement in a bilayer h-BN crystal, as shown in Fig. 2e and 2f. The large band gap (6 eV) in h-BN relative to the laser energy ($\hbar\omega\sim1.2$ eV) eliminates possible complication due to exciton-polaritons. As can be seen in Fig. 2e, while the nonlinearity of h-BN is intrinsically weaker than that in WSe\textsubscript{2}\cite{li_probing_2013}, we can still clearly identify the interference pattern in the CANS data, whereas traditional SHG under identifical experimental conditions shows nothing other than noise. The polarization dependence of the complex-valued SHG amplitude is largely three-fold symmetric but with some distortion, which is attributed to residual strain in the h-BN crystal and will be further discussed in Fig. 4. 

We now turn to the intuition about why centrosymmetric crystals do not necessarily have a vanishing SHG response. In a monolayer crystals such as monolayer 2H-TMDs and h-BN, which lacks inversion symmetry, the laser at the fundamental wavelength excites second-harmonic dipoles $\mathbf{P}^{2\omega}$, which radiates second-harmonic light waves, as shown in Fig. 3a. The reason why bilayer TMDs (or h-BN) have vanishing $\chi^{(2)}$ can be understood as a cancelation of the second-harmonic waves radiated from the top monolayer and the bottom layer. A phase difference near $\pi$ exists between these two waves because the two monolayers are $180^\circ$ rotated from each other (Fig. 3b). However, this cancelation is in fact not perfect, because the two second-harmonic waves are not completely in-phase due to the small but finite thickness $d$ of each layer ($d=$\SIrange{0.6}{0.7}{\nano\meter} for 2-H TMDs, $d=$\SI{0.34}{\nano\meter} for h-BN). We believe that the SHG wave that we measure via CANS, is the residue of such imperfect cancelation, as illustrated in Fig. 3b. 

Since a centrosymmetric system cannot possess a finite second-harmonic dipole, this residue SHG is described by the next-order multipole expansion of charge density in the crystal, which is the electric quadrupole. We shall thus refer to this wave as the quadrupolar SHG (QSHG), which was first observed in the context of isotropic gases\cite{bethune_optical_1976,bethune_quadrupole_1981}. This SHG mechanism is also responsible for the bulk SHG contribution in thicker crystals, but generally believed to negligible for atomically thin 2D crystals. As an order-of-magnitude estimation (see SI for detailed calculation), the intensity of QSHG in bilayer 2H-TMDs is $\sim10^{-4}$ weaker than that of a monolayer, and $\sim10^{-5}$ weaker for h-BN. This explains why the QSHG signal could not be readily observed via traditional SHG in these ultrathin crystals, but can now be resolved using the ultra-sensitive CANS technique.

The QSHG is universally allowed in any crystal, including those with inversion symmetry, and is fundamentally different in nature compared to the prior observation of SHG in centrosymmetric crystals where an external symmetry-breaking field was applied or induced\cite{dean_second_2009, tom_second-harmonic_1983, yao_enhanced_2021,kim_stacking_2013,hsu_second_2014,zhang_emergent_2023,an_enhanced_2013,yu_charge-induced_2015,klein_electric-field_2017,zhang_doping-induced_2019,lee_electrical_2016,ahn_electric_2024,shree_interlayer_2021,lu_strong_2023,zhang_strain-driven_2025}. While a SiO\textsubscript{2} substrate is present in our present experiments, we believe that the substrate effect is not the dominant reason why an SHG signal can be observed, and we expect that the QSHG can be observed even in fully suspended (and thus fully centrosymmetric) bilayer crystals, using CANS (see SI for estimation).

\begin{figure*}[!ht]
    \centering
    \includegraphics[width=0.95\linewidth]{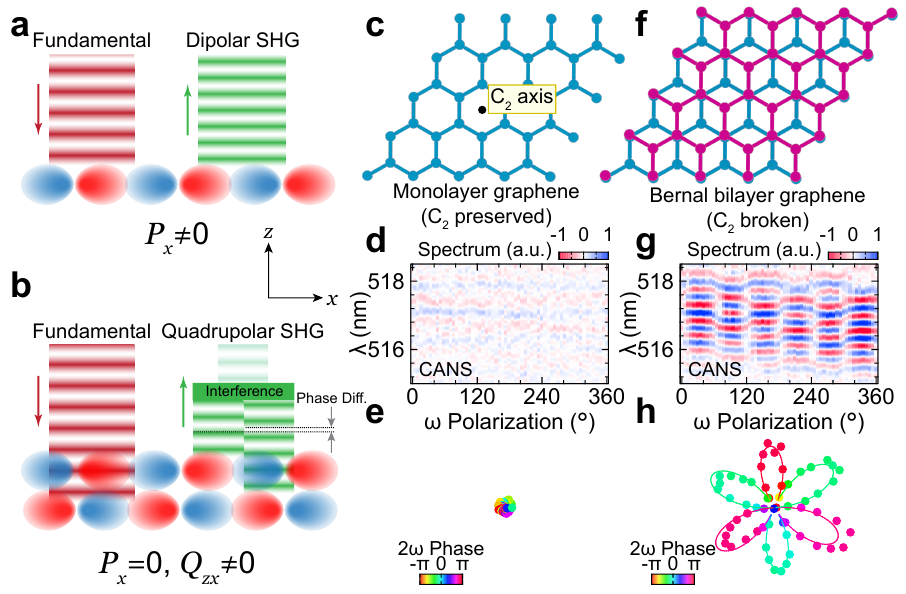}
    \caption{\textbf{Quardrupolar SHG and $C_2$ rotational symmetry.} (a) In monolayer nonlinear crystals, SHG is generated by oscillating dipoles in the plane (e.g. $P_x^{2\omega}$ along the $x$ direction), excited by the fundamental. These dipoles radiate SHG waves in the $z$ direction. (b) In a bilayer crystal (2H-TMDs or h-BN), the SHG waves from the two oppositely rotated layers cancel, but a small residue remains. The total dipole density $P_{x}^{2\omega}$ vanishes, but the quadrupole density $Q_{zx}^{2\omega}$ is nonzero, which is can still radiates in the $z$ direction, giving rise to a detectable QSHG signal. (c) and (f) shows the crystal structure of monolayer graphene (MLG) and Bernal bilayer graphene (BLG), respectively. While both are centrosymmetric, MLG has a $C_2$ rotation axis while BLG does not. (d) MLG does not show any CANS signal apart from an polarization independent background. (g) BLG shows a clear CANS signal that can be attributed to QSHG in this $C_2$-broken crystal. (e) and (h) show the polarization-resolved complex-valued SHG amplitude extracted from (d) and (g), respectively.}
    \label{fig:fig3}
\end{figure*}

\section{Quadrupolar SHG and $C_2$ symmetry}

To further demonstrate that the observed SHG in centrosymmetric crystals is from the quadrupolar contribution and to gain further insight into its symmetry properties, we performed CANS measurement on two other centrosymmetric 2D crystals, which are monolayer graphene (MLG) and Bernal bilayer graphene (BLG), respectively. 

Analogous to $\chi^{(2)}$ that describes the dipolar SHG, QSHG in centrosymmetric crystals is described by a rank-4 tensor $\chi^{Q}$ (see SI). $\chi^Q_{ijkl}$ is even under inversion and thus does not vanish in centrosymmetric crystals. However, relevant components of $\chi^Q$ could be subject to other symmetry constraints. In particular, one could show that the components in $\chi^Q$ that are responsible for radiating SHG waves into the $z$ direction must vanish when $C_2$-symmetry along the $z$-axis is present (see SI for derivation)\cite{popov_susceptibility_2017}. Therefore, whether the observed CANS signal vanishes when $C_2$-symmetry is present would be a strong indication of its quadrupolar origin.

The contrast between MLG/BLG provides an ideal testbed for this hypothesis. Both being centrosymmetric and consisting of identical atoms, MLG possesses a $C_2$ symmetry axis at the hexagonal centers, whereas BLG does not possess this symmetry\cite{wallace_band_1947}, as illustrated in Fig. 3c and f. We prepare pristine MLG and BLG by mechanical exfoliation on identical substrates. As shown in Fig. 3d and g, with identical experimental conditions, we could not detect any SHG signal in MLG (except a weak polarization-independent background from the substrate\cite{dean_second_2009}), while a strong CANS response is detected from BLG, in complete agreement with the above discussions about QSHG. Therefore, we conclude that the observed SHG in centrosymmetric $C_2$-broken crystals can be attributed to the quadrupolar contribution. 

Similar to bilayer 2H-TMDs and h-BN (Fig. 2), the complex-valued SHG amplitude of BLG extracted from the CANS data shows a $C_3$-symmetric pattern (Fig. 3h), allowing unique determination of its lattice orientation. To our knowledge, this experiment constitutes the first determination of crystallographic orientation in pristine BLG via SHG. This establishes CANS as an extremely useful tool for the preparation of moiré 2D crystals that involve the alignment between BLG and other rhombohedral graphene flakes (all of which possess inversion symmetry), with an h-BN substrate flake\cite{lu_fractional_2024}. Traditionally, the preparation of such samples involves visual alignment between the edges of h-BN and graphene flakes, which is error-prone (cannot tell between AC and ZZ edges, not to mention the ambiguity between AC+ and AC-), inaccurate (uncertainty of $2\sim3^\circ$), and irreproducible\cite{lau_reproducibility_2022}. CANS might increase the yield rate of such devices by several times if the crystallographic orientations of the underlying graphene and h-BN crystals are pinpointed before stacking.

\begin{figure*}[!ht]
    \centering
    \includegraphics[width=0.9\linewidth]{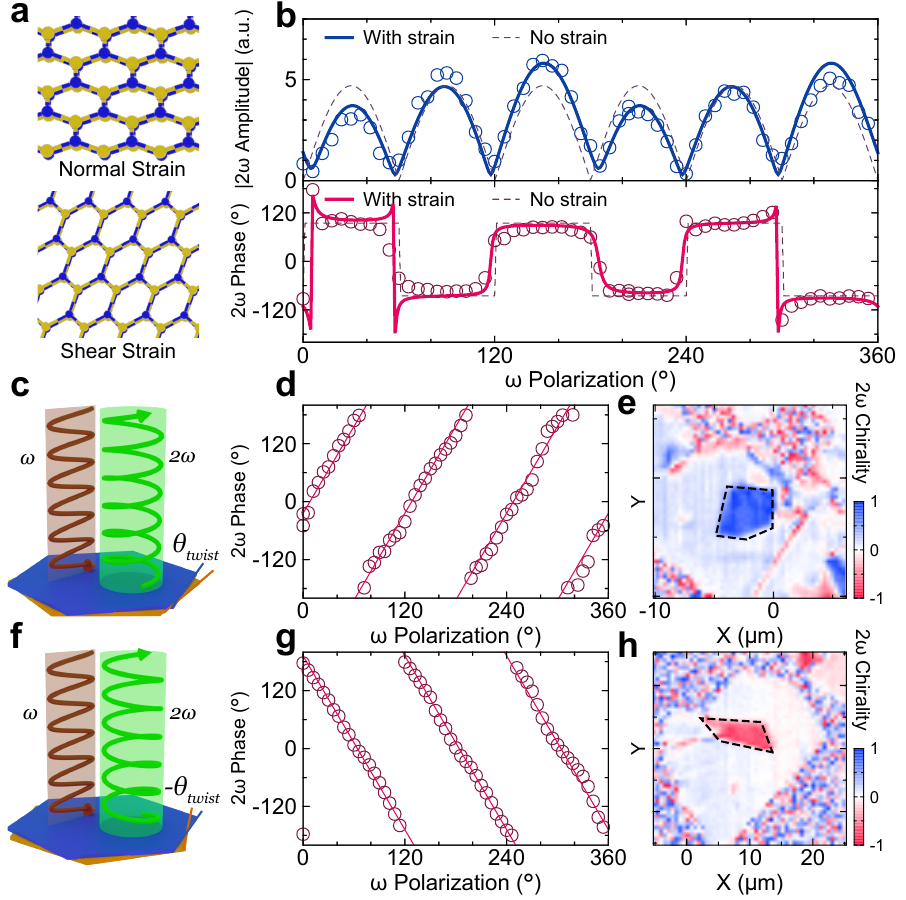}
    \caption{\textbf{Strain and geometric phase sensing using CANS}. (a) Uniaxial strain (including normal and shear strains) in a 2D crystal breaks the threefold rotational symmetry ($C_3$). (b) Complex-valued SHG amplitude and phase in a bilayer h-BN crystal with strain. The empty dots are data, solid curves are fitting that includes strain effects, and dashed curves are fitting without strain effects (see SI). (c) In a chiral system (e.g. a twisted crystal with twist angle $\theta_\mathrm{twist}$), a linearly polarized fundamental input light ($\omega$) becomes left circularly polarized (LCP) at the second harmonic frequency ($2\omega$). The Pancharatnum-Berry phase (geometric phase) is the optical phase of the $2\omega$ wave, which is related to the fundamental polarization $\theta$ by $\Delta\phi=3\theta$ when the $2\omega$ wave is purely LCP. (d) The SHG phase measured by CANS in a $\theta_{twist}=66^\circ$ pentalayer-pentalayer WS\textsubscript{2} device, where the SHG is close to LCP. The fitting curve is $3\theta$ with an offset that represents the lattice orientation. (e) SHG chirality map of the $\theta_{twist}=66^\circ$ device. The dashed line highlights the twisted region. (f-h) Same measurements for a $\theta_{twist}=57^\circ$ device, where the system has opposite chirality and the SHG is right circularly polarized (RCP). The geometric phase is $\Delta\phi=-3\theta$ plus an offset in this case. }
    \label{fig:fig4}
\end{figure*}

\section{Sensing strain and geometric phase}

The complex-valued SHG amplitude extracted from CANS contains rich information on the nonlinear crystals beyond their lattice orientation. Here we demonstrate that symmetry-breaking effects in ultrathin crystals caused by uniaxial strain or twisting can be sensed by CANS as well. CANS can thus be used as a non-contact method to measure strain and twist angle in centrosymmetric crystals that could not be sensed by traditional SHG spectroscopy.

A uniaxial strain in a 2D crystal, which could include a combination of normal strain and shear strain, breaks the $C_3$ symmetry of a hexagonal 2D crystal (Fig. 4a). The effect of such strain-induced symmetry breaking on traditional SHG has been extensively studied in noncentrosymmetric 2D crystals, which is captured by a modulation on the SHG intensity with a periodicity of $180^\circ$ (i.e. $C_2$ symmetric), on top of the standard $60^\circ$-periodic ($C_6$ symmetric) pattern\cite{mennel_optical_2018,mennel_second_2018}. However, since strain itself does not break the inversion symmetry, strain sensing using traditional SHG is not possible if the crystal is centrosymmetric. With the introduction of CANS, such sensing is now possible. Figure 4b shows the complex-valued SHG amplitude and phase for a bilayer h-BN sample (same as the one used in Fig. 2e and f). It can be seen that the data match perfectly with the fitting curves that consider strain effects, from which we can directly pinpoint the principal axes and relative strength of the uniaxial strain in this sample (see SI for further details).

Symmetry is also broken by twisting one 2D crystal against another, creating a moiré bilayer. A moiré bilayer consisting of hexgonal 2D crystal breaks both inversion symmetry and mirror symmetry, if the twist angle $\theta_{twist}$ is other than a multiple of $60^\circ$. In these crystals, one expects the nonlinear response to become chiral, \emph{i.e.} the symmetry between left and right circular optical polarizations is lifted. In particular, we have previously shown that the nonlinear response of a moiré bilayer can become fully chiral\cite{tang_-chip_2023} at specific combinations of layer thicknesses and twist angles. Under these conditions, the SHG from the moiré bilayer becomes fully left or right circularly polarized (LCP or RCP) when excited with linearly polarized fundamental, as illustrated in Fig. 4c and f. At these peculiar singularities points, the linear polarization angle of the fundamental wave, $\theta$, no longer affects the second-harmonic polarization (which is now fixated to LCP or RCP); instead, it directly controls the optical phase of the LCP or RCP light known as the Pancharatnum-Berry phase or geometric phase, given by $e^{\pm3i\theta}$ (positive for LCP and negative for RCP) for hexagonal crystals\cite{bhagavantam_harmonic_1972,li_continuous_2015,xie_generalized_2021}. 

The phase sensitivity of CANS allows us to directly visualize this geometric phase. In this experiment, we fabricated two twisted pentalayer (5L) WS\textsubscript{2} devices, one with twist angle $\theta_{tw}=66^\circ$ and one with $\theta_{tw}=57^\circ$. The experimentally measured SHG phase for these two samples are shown in Fig. 4d and g, respectively. It can be seen that in stark contrast to nonchiral crystals where the phase switches by approximately $\pm\pi$ each time the SHG amplitude drops to zero (see Fig. 2c), the CANS phase for a chiral crystal is approximately linear with the fundamental polarization angle, $\Delta\phi\sim\pm3\theta$, in precise agreement with the geometric phase for hexagonal crystals. We also performed CANS mapping in the two twisted samples to map out the SHG chirality (see SI for extraction), as shown in Fig. 4e and h respectively. The twisted regions, which exhibit close-to-unity SHG chirality, can be clearly identified in the chirality map. CANS mapping therefore allows one to remotely image the internally twisted interfaces in a crystal on the basis of its optical chirality. The same principle can be used to inspect internal twisted faults in other materials regardless of whether the underlying crystal have inversion symmetry or not.

\section{Discussion}

We have shown via several experiments that CANS is a powerful tool for optical characterization of centrosymmetric crystals where traditional SHG cannot be employed. We want to link this apparent `evading' of symmetry constraints to a more general character of perturbative physical processes in nature: while the lowest-order response of a physical process (e.g. the dipolar SHG) might vanish due to a symmetry constraint, the next-lowest-order response (e.g. the QSHG) is often not subject to the same constraint, and can thus be probed by a sufficiently sensitive experimental technique. A classical example of this philosophy is optical selection rules, which are different for electric dipole radiation and electric quadrupole radiation. Based on this philosophy and our demonstration, one can extrapolate that other physical processes typically thought of as being forbidden by the presence of inversion symmetry, such as piezoelectricity, ferroelectricity, and electrical/optical rectification, might be permitted in their corresponding quadrupolar channels, which can in principle be measured using quadrupole-sensitive techniques. 

A potentially important application of QSHG spectroscopy is to study the symmetry-broken phases in moiré and non-moiré quantum materials \cite{cao_correlated_2018,zhou_half-_2021,wong_cascade_2020,zondiner_cascade_2020}. In hexagonal 2D crystals with two valleys, the $C_2$ symmetry and the time-reversal symmetry play a central role in low-energy electronic properties\cite{castro_neto_electronic_2009}, and the theoretically predicted ground states of the correlated states in moiré 2D crystals are often classified by whether they spontaneously break one or both of these symmetries\cite{xie_nature_2020,sheffer_symmetries_2023}. However, it is difficult to non-invasively detect $C_2$ symmetry breaking and differentiate from inversion symmetry breaking with existing approaches. With the unique sensitivity to the $C_2$ symmetry, we envision that QSHG imaging could become a new approach for sensing the spontaneously $C_2$-broken phases in moiré systems.

\section*{Acknowledgements}
We thank Shanhui Fan, Alp Sipahigil, Peter So, and Pablo Jarillo-Herrero for helpful discussions. Y.C. acknowledges support by the U.S. National Science Foundation under grant No. DMR-2441893. H.T. acknowledges support by the Sony Research Award from Sony Group. F.W acknowledges support by the U.S. Department of Energy, Office of Science, Basic Energy Sciences, Materials Sciences and Engineering Division under Contract No. DE-AC02-05-CH11231 within the van der Waals heterostructure program KCWF16. Y.C. and Z.H. acknowledge support from Lawrence Berkeley National Laboratory via the LDRD program. K.W. and T.T. acknowledge support from the Elemental Strategy Initiative conducted by the MEXT, Japan, grant no. JPMXP0112101001; JSPS KAKENHI grant no. JP20H00354; and the CREST (JPMJCR15F3), JST.
\\

\section*{Author Contributions}
Y.C. and H.T. conceived and built the experiment. Z.D., T.R. and Z.H. prepared the sample and performed the measurements. K.W. and T.T. provided h-BN crystals. H.W., A.J. and F.W. provided TMDs samples. Y.C. and H.T. wrote the manuscript with inputs from all authors.

\end{document}